\documentstyle[preprint,aps,prl,epsbox]{revtex}
\tightenlines
\begin{document}
\draft
\title{Measurement of the Solar Neutrino Energy Spectrum Using
Neutrino--Electron Scattering}
%
\maketitle

\begin{center}
\newcounter{foots}

The Super-Kamiokande Collaboration \\

Y.Fukuda$^a$, T.Hayakawa$^a$, E.Ichihara$^a$, K.Inoue$^a$,
K.Ishihara$^a$, H.Ishino$^a$, Y.Itow$^a$,
T.Kajita$^a$, J.Kameda$^a$, S.Kasuga$^a$
% ...

\footnotesize \it

$^a$Institute for Cosmic Ray Research, University of Tokyo, Tanashi,
Tokyo 188-8502, Japan\\
% ...
\end{center}

\begin{abstract}

   A measurement of the energy spectrum of recoil electrons from solar
neutrino scattering in the Super--Kamiokande detector is presented.
  The results shown here are obtained from 504 days of data taken 
between the 31$^{st}$ of May, 1996 and the 25$^{th}$ of March, 1998. 
  The shape of the  measured spectrum is compared with the expectation for
solar $^{8}$B neutrinos.  
  The comparison takes into account both kinematic and detector related 
effects in the measurement process.
  The spectral shape comparison between the observation and the expectation 
gives a $\chi^{2}$ of 25.3 with 15 degrees of
freedom, corresponding to a 4.6~\% confidence level.

\end{abstract}


\narrowtext

  Previous solar neutrino experiments \cite{Cl37,KAMALL,Sage,Gallex,SKFLUX} 
have measured significantly smaller neutrino flux than the
expectation from standard solar models 
(SSMs)\cite{SSM-BP98,SSM-BP95,SSM-TCL,SSM-TC98}, an enigma that has been
known as ``the solar neutrino problem'' for almost three decades.
Detailed studies of this discrepancy between the observations and 
the predictions strongly suggest that the apparent deficits in the 
measured fluxes are not easily explained by modifying the solar models,
but can be naturally explained by neutrino oscillations\cite{OSC}.
Since the expected spectral shape of the solar neutrinos 
can be calculated using well established results from the
terrestrial experiments, measurement of the solar neutrino
energy spectrum can provide a direct, solar model-independent test of the
neutrino oscillation hypothesis.
$^8$B solar neutrinos are detected in the Super--Kamiokande detector
by observing recoil
electrons resulting from neutrino-electron scattering in the water.
  The observed energy spectrum of recoil electrons reflects that of 
the $^8$B solar neutrinos arriving at Earth.  

In a previous letter, we reported a measurement that confirmed the solar 
$^8$B  neutrino flux deficit utilizing the
first 300 days of data \cite{SKFLUX}.
  The updated measured flux using 504 days of data is 
2.44$\pm$0.05(stat.)$^{+0.09}_{-0.07}$(syst.) $\times$10$^6$/cm$^2$/s, 
which corresponds to a ratio Data/SSM of 
0.474 $^{+0.010}_{-0.009}$ $^{+0.017}_{-0.014}$, 
using the latest calculation 
by Bahcall et al. (BP98) \cite{SSM-BP98}, and 
0.506 $^{+0.011}_{-0.010}$ $^{+0.018}_{-0.015}$, using Brun 
et al.\cite{SSM-TC98}.
In this letter we present a measurement of the recoil electron 
energy spectrum based upon 504 live days of data collected with the 
Super--Kamiokande detector.

  Super--Kamiokande, a 50,000 ton imaging water Cherenkov
detector, utilizes a 22,500 ton fiducial volume for the solar neutrino
analysis; details of the detector are described in Ref.\cite{SKFLUX}.
  The vertex position and direction of the recoil electrons are reconstructed 
by using the timing information and ring pattern of the hit photomultiplier 
tubes (PMTs)\cite{SKFLUX}.
  Vertex position and angular resolutions for 10~MeV electrons are 0.71~m 
and 26.7$^\circ$, respectively.
  Electron energy is measured by calculating the effective number of hit PMTs, 
$N_{eff}$, which is the number of hit PMTs with corrections for 
light attenuation through the water, the angular dependence of
PMT acceptance, the effective density of PMTs, the number of nonfunctioning 
PMTs, and the probability of a two photo-electron emission in one PMT.
  Further corrections are made for noise hits due to
the PMT dark noise rate ($\sim$3.3~kHz, which contributes about 1.8 hits
within 50 ns) and for the tail of the hit PMT time distribution (up to 
100 ns), caused by the scattering of light in the water and by reflections 
on the PMT and light barrier surfaces.  
  The $N_{eff}$  corrections are designed to remove position and 
water transparency-related effects so as to give uniform response over the 
fiducial volume.
  The non-uniformity of $N_{eff}$ within the detector volume is measured to
be less than 2~\% using mono-energetic electrons and gamma-rays and is 
consistent with the corresponding Monte Carlo (MC) simulations to within
0.5~\%.
  $N_{eff}$, as above described, is closely related to electron visible
energy. 
  However, the energy used in this analysis includes the energy 
deposition below the Cherenkov threshold and the rest mass of the electron 
and is, therefore, the total electron energy.  
  Thus, any difference between the measured total 
energy obtained by $N_{eff}$ and the true electron total energy is due 
to detector energy resolution smearing and position dependent response.

 Precision energy calibration of the detector is essential for the energy
spectrum measurement of recoil electrons.
We employ an energy calibration procedure using an 
electron linear accelerator (LINAC) to 
relate $N_{eff}$ to absolute energy.
  The absolute energy scale is monitored for stability and cross-checked 
using:
(1) muon decay electrons, 
(2) spallation products induced by cosmic ray muons,
(3) $^{16}$N produced by stopping muon capture on oxygen, and 
(4) a Ni(n,$\gamma$)Ni source.
  
\begin{figure}
\center
\psbox[height=12.0cm]{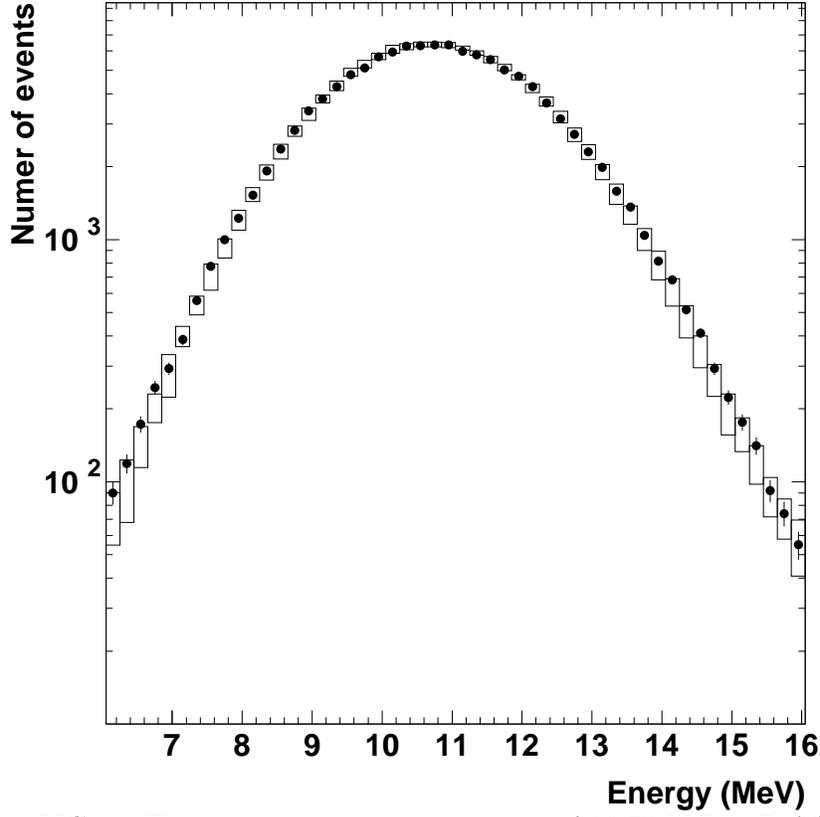}
\caption{The measured energy spectrum of 10.78~MeV LINAC electrons is shown 
by the data points. 
The data points are the sum of the values taken at 8 representative positions 
within
the detector.  The boxes are the summation of values from the corresponding MC 
simulations, where the vertical size of a box indicates the estimated 
systematic errors in energy scale and resolution added in quadrature 
with statistical error.}
\label{fig:linac}
\end{figure}

  The LINAC is used for calibrating the absolute energy scale and also
for measuring the angular and vertex position resolutions.
   Details of the LINAC calibration are described in \cite{linac},
but a brief  summary is given here.
   The LINAC, located near the Super--Kamiokande detector,
injects downward going monoenergetic single electrons into the detector tank
with a tunable energy
ranging from 5 to 16~MeV.
   The absolute energy of the beam is measured by a germanium detector, 
which was in turn calibrated by gamma-ray sources and internal-conversion 
electrons from a $^{207}$Bi source.  
   The uncertainty in the  beam energy deposition in the Super--Kamiokande
detector is 0.55~\% at 6~MeV and 0.3~\% at 10~MeV, resulting from 
the uncertainty in the beam energy ($<$ 20~keV) and the reflectivity of the  
beam pipe end-cap materials.
  Energy calibration utilizes LINAC data taken at 8 representative positions 
within the Super--Kamiokande fiducial volume with 7 different momenta 
ranging from 4.89 to 16.09~MeV.
  The absolute energy scale, the relation between $N_{eff}$ and 
the total electron energy, is obtained from a MC simulation program for
which various parameters are tuned to reproduce the LINAC data taken at 
the various positions and energies.
  The MC simulation is based on GEANT 3.21 with the
water attenuation lengths (absorption and scattering lengths) and reflectivity
of detector materials, such as the light barrier surfaces separating
the inner and outer detectors and the surfaces of 50~cm PMTs, 
as tunable parameters.
  After tuning, the MC reproduces the position dependence of the energy scale 
as measured by the LINAC to within 0.5~\% on average. 
  The energy resolution for electrons is also calibrated by the LINAC, and
the difference between LINAC data and the corresponding MC simulation
is less than 2~\%.
  Figure~\ref{fig:linac} shows the measured energy spectrum of 
LINAC 10.78~MeV data
compared with the corresponding MC simulation.
  There is good agreement in the shape over two orders of magnitude, 
demonstrating that the MC simulation accurately translates input
electron energy into energy measured by the detector.

 The large number of muon decay electrons, $\sim$1500 events/day,
allows monitoring of the temporal variation of water attenuation length.
  The measured water attenuation length is used in calculating 
$N_{eff}$ for each event to correct for the Cherenkov photon attenuation. 
The variation of the water attenuation length has caused $\sim$3.8~\% 
change in the energy scale over the data taking period considered
in this paper.
  After correcting $N_{eff}$ for the variations in water attenuation length, 
the stability of the energy scale is better than 0.5~\% over the time period
described here and $\sim$0.2~\% in r.m.s.
  This variation is included in the uncertainty in the
absolute energy scale, since the energy scale set by the
LINAC calibration is extrapolated to the entire time period of this
analysis.

The time variation and directional dependence of the energy scale was 
cross-checked using spallation events, which are beta- and gamma-rays from 
radioactive nuclei created by cosmic ray interactions within the detector.  
 Because spallation events are distributed uniformly in time and throughout the
detector volume, they can be used to monitor the time variation and the
directional dependence of the energy scale on a more continuous basis and 
at more points in the volume than is possible with the LINAC.
The resulting time variation of the energy scale checked with the 
spallation events is less than 0.5~\% over the entire time period,
consistent with the result obtained using the muon decay electron sample.
  The spallation events are subdivided into 10 data sets according to
the reconstructed zenith angle and the relative difference of the
energy distribution among the 10 data sets is compared.
  The obtained angular dependence of the energy scale is less than
0.5~\%.
  This result allows the use of the LINAC absolute energy
calibration, which thus far has been taken electrons moving 
only in the downward-going direction, for all directions.
   
Another cross-check on the absolute energy calibration is made using the
decays of $^{16}$N produced by stopping muon capture on oxygen.
These events with well defined decay lines are also uniformly
distributed in time and position,
thus, providing another natural handle on absolute energy calibration.
  The difference in energy scales between that obtained by $^{16}$N decay
beta spectrum and the MC tuned to LINAC data is 0.2 $^{+0.6}_{-0.8}$~\%.

 Summing all possible sources of the uncertainty in the absolute energy
scale described above in quadrature, the overall 
uncertainty in the energy scale is
estimated to be $\pm$0.8~\% at 10~MeV, which includes contributions from the
uncertainty in the LINAC electron energy deposition ($\pm$0.3~\%),
the position dependence of the energy scale ($\pm$0.5~\%), the uncertainty
of the water transparency determination ($\pm$0.2~\%), and the directional 
dependence of the energy scale ($\pm$0.5~\%).

  The energy dependence of the angular resolution of the detector is
measured by using LINAC data\cite{linac}.
  The measured angular resolution is 2--3~\% smaller than 
the corresponding MC simulation.
  The difference could be due to an inaccurate description of light scattering
in the current MC simulation, but it is not yet fully understood.
  This difference in the angular resolution is corrected for
in the solar neutrino flux calculation which follows.

For the energy spectrum measurement analysis, we follow the same data
reduction procedure described in Ref.\cite{SKFLUX}. We have used
the data obtained from 504 live days between 31 May 1996 and 25 March 1998.
  The data set (initially consisting of $\sim$7$\times$10$^8$ events)
was reduced by requiring that event vertices be within the fiducial
volume and by instituting cuts designed to reject external
gamma ray and muon-induced spallation events.
  Details of the data reduction are described in Ref.\cite{SKFLUX}. 
The total efficiency of the data reduction is
70.0~\% with an estimated uncertainty less than 0.7~\%.

\begin{figure}
\center
\psbox[height=12.0cm]{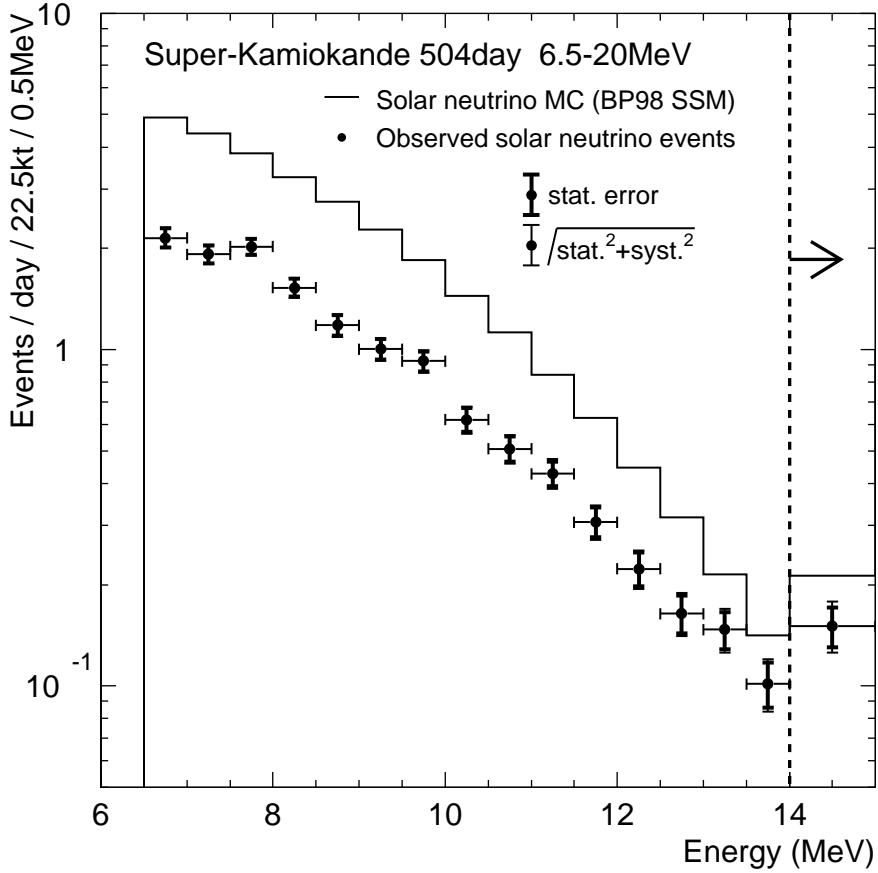}
\caption{Recoil electron energy spectrum of solar neutrinos (data points).
The histogram shows the expectation from the SSM.
The inner and outer error bars show the statistical and systematic errors,
respectively.
The systematic error is the sum of correlated experimental and calculation
errors and uncorrelated errors added quadratically.}
\label{fig:spec}
\end{figure}

\begin{figure}
\center
\psbox[height=12.0cm]{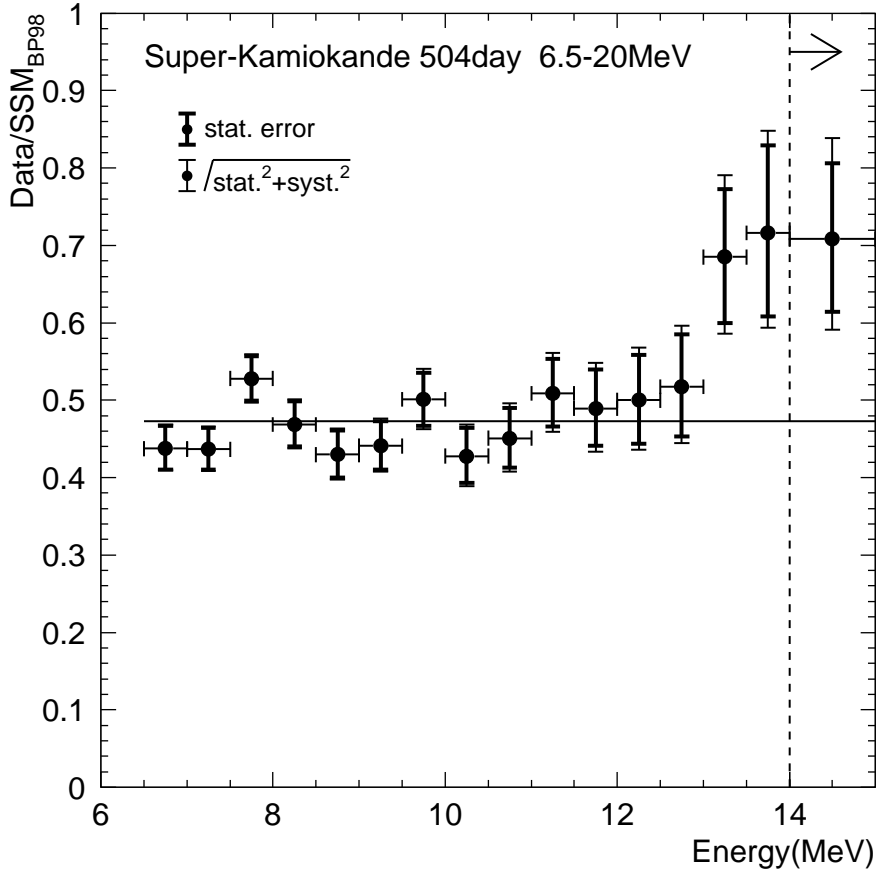}
\caption{Ratio of observed electron energy spectrum and expectation from the
SSM.
Errors are the same as in Fig. 2.}
\label{fig:ratio}
\end{figure}

   The final data sample is sub-divided into 16 energy bins, every
0.5~MeV  from 6.5 to 14.0~MeV and one bin combining events with energies from 
14.0 to 20.0~MeV.
  The number of solar neutrino events in each energy bin is extracted 
individually by analyzing the angular distribution of the events within each
bin with respect to the sun.
  The angular distribution in the region far from the solar direction 
provides a measure of the background level.
A near-isotropic background angular distribution with respect to the 
direction of the sun is obtained with small corrections made for slight 
directional anisotropies in local detector coordinates.
  The background fit along with the expected angular distributions of the
solar neutrino signal are incorporated into a maximum likelihood method to
extract the number of solar neutrino events.
  The error in the number of solar neutrino events due to possible 
local detector anisotropies using this technique is ~0.1~\%.   
  The number of solar neutrino events thus obtained is shown
in Fig.~\ref{fig:spec}.
  The measured spectrum is then compared with the expected spectrum
obtained from our MC simulation.
The MC events are generated using: 
(1) the total $^8$B solar neutrino flux from 
Ref.\cite{SSM-BP98} (5.15$\times$10$^6$/cm$^2$/s; a particular SSM is 
not required for the spectral shape analysis),
(2) the calculation of $^8$B neutrino spectral shape from
Ref.\cite{SSM-SPEC}, and (3) the electron spectrum of $\nu$-e scattering 
from Ref.\cite{NUE-CROSS}, in which radiative corrections are taken 
into account.
  The smearing of the expected recoil electron energy spectrum, mainly by the 
finite energy resolution of the detector, is done by a full detector 
simulation.
The simulated MC events are then passed through the 
same analysis chain as the data
resulting in a MC recoil energy spectrum shown as a histogram in 
Fig.~\ref{fig:spec}.
  In order to compare the shape of the observed energy spectrum with the
expectation, the ratio of observed and expected numbers of events for each
energy bin is taken; these ratios are plotted in Fig.~\ref{fig:ratio}.

  Systematic errors in the energy shape comparison are classified into
three categories:  (1) energy-bin-correlated  experimental errors (called 
``correlated'' from now on), (2) energy-bin-correlated error in the 
expected energy spectrum calculation, and (3) energy-bin-uncorrelated 
(``uncorrelated'') errors.
  The sources of correlated experimental errors are uncertainties in the
absolute energy scale and energy resolution.
  The systematic error of the electron energy spectrum due to the
correlated experimental uncertainties is shown in
Table~\ref{tab:spec}.
  For example, the systematic error of the 13.0 to 13.5~MeV energy bin
is $^{+6.9}_{-6.2}$~\%, in which $^{+6.6}_{-5.9}$~\% comes from
the uncertainty of the absolute energy scale and $^{+2.2}_{-1.9}$~\% from
the uncertainty of the energy resolution.
  The correlated error in the expected spectrum calculation is 
obtained by using the 1$\sigma$ error of $^8$B neutrino energy spectrum 
described in Ref.\cite{SSM-SPEC} and shown in
Table~\ref{tab:spec}.
  The sources of uncorrelated errors are: the uncertainty in trigger efficiency
(+1.2~\% error in energy spectrum only for 6.5--7.0~MeV energy bin), 
the uncertainty in the data reduction efficiency ($\pm$0.7~\%),
the uncertainty in the directional anisotropy of the background
($\pm$0.1~\%), and the uncertainty in the size of the fiducial volume by
possible systematic
shift of the vertex position ($\pm$1.0~\%).
  Uncertainties which may be energy-bin-correlated, but whose energy dependence
is not well known, are categorized as uncorrelated systematic errors
by assigning the largest possible deviation in the energy spectrum to
each energy bin.
  Such errors include the uncertainty in angular resolution ($\pm$1.0~\%) and
the uncertainty in the cross section of $\nu$-e scattering ($\pm$0.5~\%).
The sum of uncorrelated errors is shown in Table~\ref{tab:spec}.

The observed energy spectrum is examined using the
following $\chi^{2}$ :
\begin{eqnarray}
\chi^2 = \sum_{i=1}^{16} \{
\frac{ ( \frac{data}{SSM} )_i -
\alpha /
(( 1 + \delta_{i,exp} \times \beta )
( 1 + \delta_{i,cal} \times \gamma ) )}
{\sigma_i} \}^2 + \beta^2 + \gamma^2,
\nonumber
\end{eqnarray}
where $\delta_{i,exp}$ and $\delta_{i,cal}$ are 1$\sigma$ errors of
the correlated experimental error and of the expected spectrum calculation
described above, $\sigma_i$ is a 1$\sigma$ error for each
energy bin defined as
a sum of statistical error and uncorrelated errors added quadratically, and
$\alpha$ is a free parameter which normalizes the measured 
$^8$B solar neutrino flux  relative to the expected flux.
  $\beta$ and $\gamma$ are also free parameters used for constraining the
variation of correlated systematic errors.
  The minimum value of this $\chi^2$ is obtained by numerically 
varying the free parameters, which results in a minimum value of 25.3 
(with 15 degrees of freedom), a value of $\alpha$ of 0.449,  and values 
of $\beta$ and $\gamma$ (measured in standard deviations) of $-1.49$ and
$-0.93$, respectively. 
The resulting minimum  $\chi^2$ corresponds to an agreement of the 
measured energy shape with the expected energy shape at the 4.6~\%
confidence level.
  The rather poor fit of  $\chi^2$ is due mainly to the rise in the 
observed spectrum at the high energy end, where the uncertainties in the
absolute energy scale and resolution can have large effects.
  To account for the rise with these uncertainties, the absolute energy scale 
must be shifted, horizontally, by 3.6~\% or the energy resolution worsened by 20~\%.
These values are 4 and 10 times larger than the estimated uncertainties,
respectively. Hence, the rise in the spectrum is difficult to be 
explained by these experimental uncertainties.

The contribution of high energy solar neutrinos from
$^3He + p \rightarrow ^4He + e^+ + \nu_e$ (hep) is estimated using the
best estimate flux in the SSM\cite{SSM-BP98}, 2.1$\times$10$^3$/cm$^2$/s .
  The expected number of events during the 504 live day period is
0.84 and 1.3 events for energy ranges of 13--14~MeV and 14--20~MeV, 
respectively.
  The uncertainty in the SSM calculation of the hep neutrino
flux is not precisely known, but an explanation of the high-energy
points in terms of a hep neutrino component would require a dramatic
increase in the SSM expectation. 
 
In conclusion, we have measured recoil electron energy spectrum 
from $^8$B solar neutrinos with the Super--Kamiokande detector. 
  A comparison of the observed spectrum with the expectation exhibits 
a poor agreement at the 4.6~\% confidence level.
  An interpretation in terms of neutrino oscillations  will be published later.
\\
\\
   We gratefully acknowledge the cooperation of the Kamioka Mining and Smelting
Company. 
   This work was partly supported by the Japanese Ministry of Education,
Science and Culture and the U.S. Department of Energy.

\newpage
\begin{table}
\begin{center}
\begin{tabular}[h]{|r|r|r|r|}  \hline
energy &  $\delta_{i,exp}$ &
$\delta_{i,cal}$ & $\delta_{i,uncorrelated}$ \\
(MeV) & & &  \\ \hline
6.5-7.0 & 
$^{+1.3}_{-1.2}$\% & $^{+0.5}_{-0.1}$\% & $^{+2.1}_{-1.7}$\% \\
7.0-7.5 &
$\pm$1.3\% & $^{+0.6}_{-0.3}$\% & $\pm$1.7\% \\
7.5-8.0 & 
$\pm$1.5\% & $^{+0.8}_{-0.5}$\% & $\pm$1.7\% \\
8.0-8.5 & 
$\pm$1.8\% & $^{+1.0}_{-0.7}$\% & $\pm$1.7\% \\
8.5-9.0 & 
$^{+2.1}_{-2.2}$\% & $^{+1.2}_{-1.0}$\% & $\pm$1.7\% \\
9.0-9.5 & 
$\pm$2.5\% & $^{+1.5}_{-1.2}$\% & $\pm$1.7\% \\
9.5-10.0 &
$\pm$2.9\% & $^{+1.8}_{-1.5}$\% & $\pm$1.7\% \\
10.0-10.5 &
$\pm$3.3\% & $^{+2.1}_{-1.8}$\% & $\pm$1.7\% \\
10.5-11.0 & 
$^{+3.8}_{-3.7}$\% & $^{+2.4}_{-2.1}$\% & $\pm$1.7\% \\
11.0-11.5 & 
$^{+4.3}_{-4.2}$\% & $^{+2.8}_{-2.4}$\% & $\pm$1.7\% \\
11.5-12.0 & 
$^{+4.9}_{-4.6}$\% & $^{+3.2}_{-2.7}$\% & $\pm$1.7\% \\
12.0-12.5 &
$^{+5.5}_{-5.1}$\% & $^{+3.7}_{-3.0}$\% & $\pm$1.7\% \\
12.5-13.0 &
$^{+6.2}_{-5.7}$\% & $^{+4.2}_{-3.4}$\% & $\pm$1.7\% \\
13.0-13.5 & 
$^{+6.9}_{-6.2}$\% & $^{+4.7}_{-3.8}$\% & $\pm$1.7\% \\
13.5-14.0 & 
$^{+7.7}_{-6.8}$\% & $^{+5.2}_{-4.1}$\% & $\pm$1.7\% \\
14.0-20.0 &
$^{+9.9}_{-8.5}$\% & $^{+6.7}_{-5.2}$\% & $\pm$1.7\% \\
\hline
\end{tabular}
\end{center}
\caption{
1$\sigma$ error of the flux due to correlated experimental error
(2nd column), due to calculation of the expected spectrum (3rd column),
and due to uncorrelated systematic error (4th column).}
\label{tab:spec}
\end{table}

\newpage
\newpage
\newpage
%

\end{document}